# Learning to Detect and Mitigate Cross-layer Attacks in Wireless Networks: Framework and Applications

Liyang Zhang, Francesco Restuccia, and Tommaso Melodia
Department of Electrical and Computer Engineering
Northeastern University
Boston, MA 02215 USA
Email: {liyangzh, frestuc, melodia}@northeastern.edu

Scott M. Pudlewski
Air Force Research Laboratory
RITF
Rome, NY 13440 USA
Email: scott.pudlewski.1@us.af.mil

*Abstract*—**Security threats such as jamming and route manipulation can have significant consequences on the performance of modern wireless networks. To increase the efficacy and stealthiness of such threats, a number of extremely challenging, cross-layer attacks have been recently unveiled. Although existing research has thoroughly addressed many single-layer attacks, the problem of detecting and mitigating cross-layer attacks still remains unsolved. For this reason, in this paper we propose a novel framework to analyze and address cross-layer attacks in wireless networks. Specifically, our framework consists of a *detection* and a *mitigation* component. The attack detection component is based on a Bayesian learning detection scheme that constructs a model of observed evidence to identify stealthy attack activities. The mitigation component comprises a scheme that achieves the desired trade-off between security and performance. We specialize and evaluate the proposed framework by considering a specific cross-layer attack that uses jamming as an auxiliary tool to achieve route manipulation. Simulations and experimental results obtained with a test-bed made up by USRP software-defined radios demonstrate the effectiveness of the proposed methodology.**

## I. Introduction

It is widely known that wireless networks are vulnerable to various types of attacks at all layers in the protocol stack [1]. Traditionally, researchers have studied each of these attacks *individually*, and have correspondingly devised dedicated countermeasures [2], [3], [4]. While these approaches have resulted in effective defense from individual attacks, they may not be directly applicable to emerging *cross-layer* attacks [5], [6], [7], which have *activities* and *objectives* that entail different layers of the network protocol stack.

**Motivations and Examples.** To improve the performance of wireless networks, functionalities such as power allocation, channel selection, routing decision, are often jointly optimized at different layers with a common optimization objective [8], [9], [10]. This provides potential adversaries with an alternative (and effective) attack strategy. Specifically, instead of attacking the target layer directly, the adversary may choose to attack a different layer. In this way, small-scale attack activities on this layer may lead to dramatic changes on the target layer. Emerging software-defined radio platforms further facilitate such attacks by easing the manipulation of lower-layer dynamics.

Cross-layer attacks present several advantages compared to traditional single-layer attacks. To illustrate this point, we discuss some examples.

• *MAC Poisoning.* Let us suppose a node has two frequency channels ($f_0$ and $f_1$) dynamically available for communication. Let us also suppose $f_0$ is experiencing more interference than $f_1$, thus the node is using $f_1$. The target of the adversary is decreasing the node's throughput. In case of a single-layer attack, the attacker may directly jam the physical layer on $f_1$. However, in case of cross-layer attack, the attacker may influence the medium access control (MAC) layer by periodically falsifying channel reservation on frequency $f_1$, and thus inducing the node to eventually switch to $f_0$. This will lead the node to ultimately experience lower throughput.

• *Hammer-and-Anvil [5].* In this attack, the considered scenario is a wireless multi-hop network as illustrated in Figure 1. The objective of node $n$ is to minimize the expected end-to-end delay from itself to the sink.

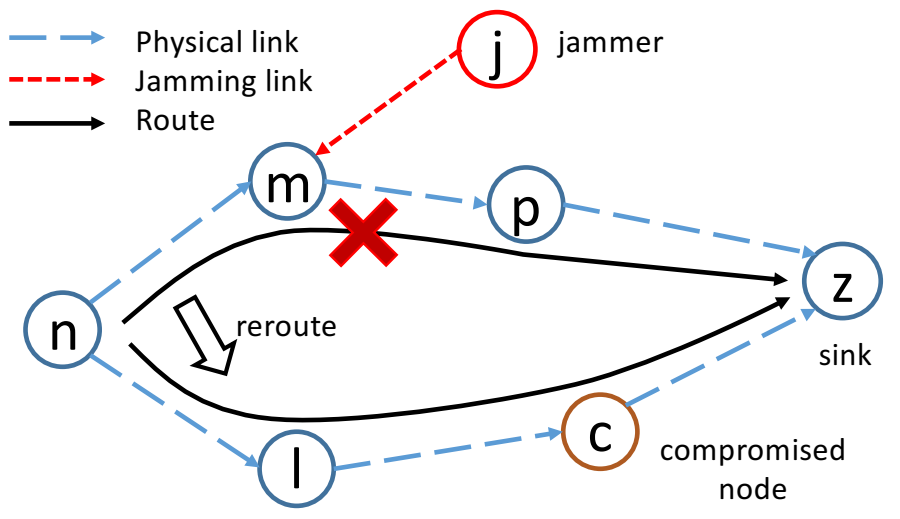

Fig. 1. Illustration of hammer-and-anvil attack.

The attack takes place by using a jammer and a compromised node, which we assume is able to undermine communication with selective forwarding, traffic analysis, or decryption, among others. The jammer aims at redirecting traffic to the compromised node. To this end, it selectively jams links that do not lead to the compromised node. With appropriate jamming, the link between $n$ and $m$ will be degraded so much that $m$ is no longer the best next hop for $n$, Thus, $n$ may select another next hop with a higher utility, which also leads to the compromised node in this example.

• *TCP Timeout.* In this attack, the adversary aims at stealthily disrupting existing TCP flows of a node. Similar to [11], a compromised node in the path can delay the forwarding of packets to the destination, thus increasing the round-trip time (RTT) of packets and ultimately leading the TCP sender to increase the timeout at each retransmission. Specifically, if the delay in forwarding is carefully chosen and sufficient to induce enough packet losses, the TCP flow will enter a timeout and attempt to send a new packet retransmission timeout (RTO) seconds later. If the period of the attack approximates the RTO of the TCP flow, the TCP flow will continually incur loss as it tries to exit the timeout state, fail to exit timeout, and obtain near zero throughput.

**Challenges.** The examples above highlight the peculiar characteristics of cross-layer attacks. First, by properly selecting a layer to attack, significant impact can be achieved even with small-scale attack activities, making well-conceived cross-layer attacks typically harder to detect than single-layer attacks. Second, a cross-layer attack may induce the victim to believe that a certain choice on a particular layer is optimal, while in fact it also increases the risk of the system. For example, in Hammer-and-Anvil, the victim reroutes its traffic to another node to achieve better performance, but this in turn results in an increased control over the data by the attacker.

These characteristics pose significant challenges to both detection and mitigation of cross-layer attacks, which existing single-layer solutions fail to address. For example, to address jamming at the physical layer, methods based on an expected packet delivery ratio (PDR) are widely used [12], [13], [14]. However, these approaches may not be valid to tackle cross-layer attacks, since jamming may only be an auxiliary tool to disrupt the functionality of another layer. Thus, relying on expected PDR (or other threshold-based criteria) will result in a low attack detection rate. This calls for a detection scheme capable of detecting small-scale activities. As far as mitigation is concerned, single-layer solutions aim at evading from the attacked state, which also fails in the context of cross-layer attacks. In fact, a very popular anti-jamming method is to exploit redundancy in routing [15], [16], which is the basis of the Hammer-and-Anvil attack. Indeed, a good mitigation scheme must be able to achieve a desirable performance-security trade-off, according to different applications.

**Novel Contributions.** To address these problems, in this paper we design and develop a framework to ensure high-throughput and reliable wireless networking in the presence of cross-layer attacks. To effectively detect stealthy attacks, we design a learning scheme based on Bayesian learning [17] which cumulatively uses weak evidence to identify stealthy attack activities. We also design a mitigation component to optimize the trade-off between security and performance. The detection and mitigation schemes work together to create a countermeasure to cross-layer attacks.

To evaluate the framework on a practical use-case scenario, we apply the framework to solve the hammer-and-anvil attack [5]. Experimental results obtained with simulations and a practical test-bed implemented with Ettus USRP software-defined radios [18] show that our framework is able to tackle such attack effectively, even when the jamming activities are performed at a level of 10x the background noise level.

**Paper Organization.** The rest of the paper is organized as follows. Section II formally introduces the proposed framework, while a use-case scenario of our framework is considered in Section III. Simulation and experimental results are introduced in Section IV. Related work is discussed in Section V, while Section VI draws conclusions.

## II. Detection and Mitigation Framework

The objective of the framework is to enable legitimate users in a wireless network to detect and mitigate possible cross-layer attacks. An overview is shown in Figure 2.

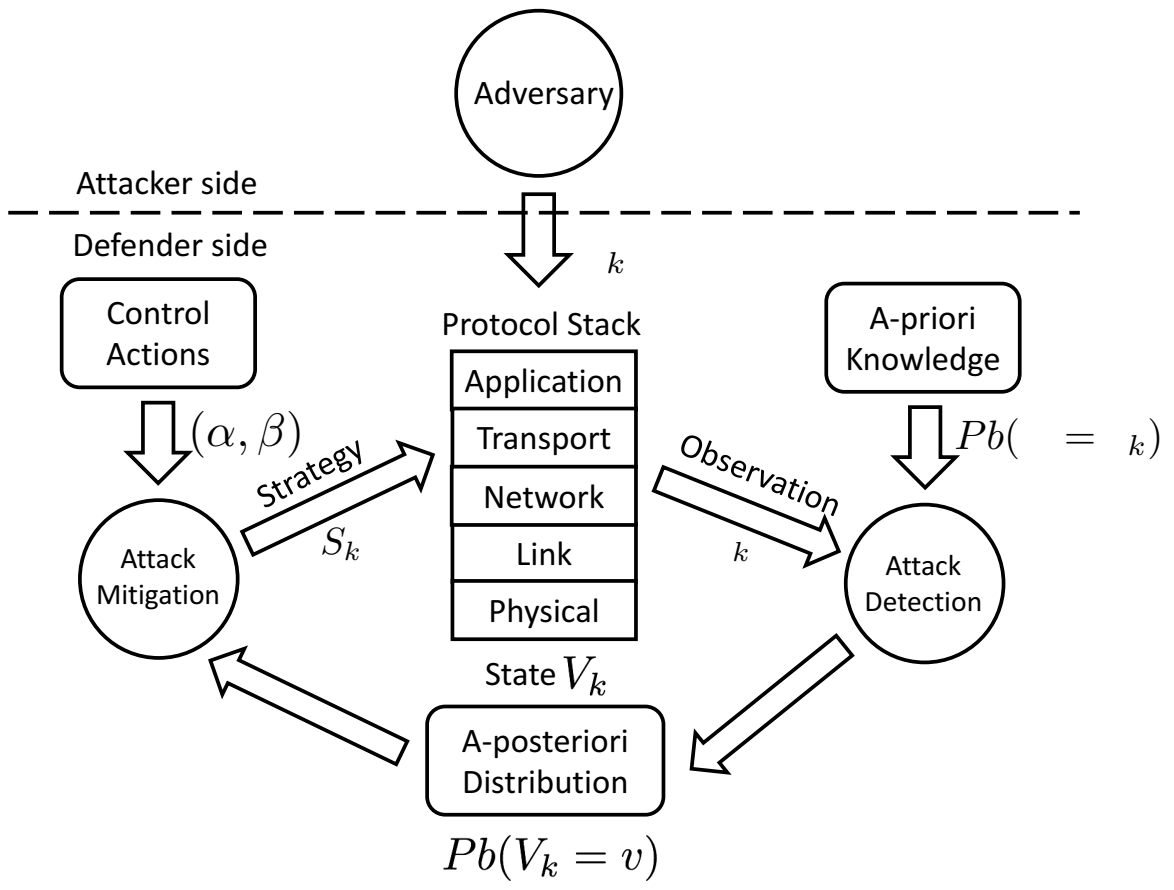

Fig. 2. Attack detection and mitigation framework.

### A. Attack Model

We consider a wireless network with a set of legitimate users and an adversary, which will be referred to as the *defenders* and the *attacker*. Defenders and attackers interact with the network and influence its status with their strategies.

The *network state* can be fully described by a set of variables on multiple layers, which may include signal-to-interference-plus-noise ratio (SINR) of a link, the channel access probability of a node, the quality of a route, among others. We define $\mathbf{V}$ as the vector holding the state.

$\mathbf{V}$ is affected by strategies of the defenders and the attacker, which may include power allocation, scheduling, and next hop selection. Denoting the strategies for a defender and an attacker as $\mathbf{S}$ and $\mathbf{A}$, respectively, then $\mathbf{V}$ can be expressed as

$$\mathbf{V} = g(\mathbf{S}, \mathbf{A}), \tag{1}$$

for some function $g$.

The attacker and the defenders update their strategies in an iterative way. The attacker updates its strategy only when at least one targeted defender has updated its own, and we define the interval between two updates of the attacker's strategy as one *strategy updating period*. The strategy updating iteration is shown in Figure 3, where subscripts such as $i$ and $i+1$ are used to distinguish different periods.

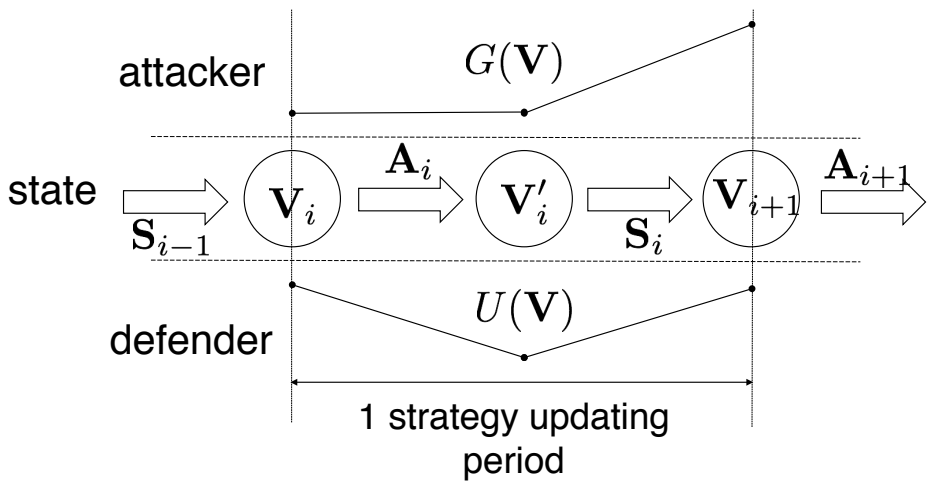

Fig. 3. Attack-defend iteration.

The defender complies with a cross-layer protocol, in which variables on multiple layers are jointly optimized, with a *utility* function $U(\mathbf{V})$. The strategy is decided by solving

$$\mathbf{S}_{i+1} = \arg\max_{\mathbf{S}} U(\mathbf{V}_i) = \arg\max_{\mathbf{S}} U(g(\mathbf{S}, \mathbf{A}_i)). \quad (2)$$

The attacker launches a cross-layer attack, with an objective unknown a-priori to the defender. Different rewards are obtained depending on the network state, which will be referred to as the *attack gain* and denoted by $G(\mathbf{V})$. To launch the attack, the attacker chooses a layer to perform attack activities and create small-scale changes on. Therefore, a network state $\mathbf{V}'_i$ that is only slightly different from $\mathbf{V}_i$, is obtained. For the aimed defender, a lower utility $U(\mathbf{V}'_i)$ is experienced, and it will seek a new optimal solution according to (2). By choosing a proper layer to attack, it is possible that the optimization path for $U(\mathbf{V})$, $\mathbf{V}'_i \to \mathbf{V}_{i+1}$ also increases $G(\mathbf{V})$. In other words, in response to the small change made by the attacker on one layer, the defender changes the states on other layers, in a direction that both maximizes its utility and increases the attack gain.

### B. Attack Detection

The detection engine computes the belief that a specific cross-layer attack is taking place. First, it estimates the attack activities based on observed events; then, the result is mapped to possible attacks according to their features.

The attack activities are directly reflected by the network state changed by the attacker. However, since $\mathbf{V}'$ is only slightly different from $\mathbf{V}$ (we omit the subscripts such as $i$ and $i+1$ since the following discussion only involves one strategy updating period), the defender must be able to detect slight changes. To address this challenge, we use Bayesian learning [17], which treats the unknown as a random variable and update its distribution once a new evidence is available. The more evidences are accumulated, the closer the result is to the real distribution of the variable. Therefore, the defender is able to construct a hypothesis with high confidence based on consecutive but not decisive evidences, which perfectly suits the scenario we are considering.

Specifically, $\mathbf{V}$ is viewed as a random variable with an unknown distribution. This assumption is valid, since the strategy of the attacker is static in one strategy updating period. Note that some environmental variables may also affect $\mathbf{V}$, but they can be generally assumed to be either constant (e.g., network topology), or a random variable with a static distribution (e.g., channel fading) during one strategy updating period. Therefore, these variables do not undermine the assumption.

To learn the true distribution of $\mathbf{V}$, the defender keeps records on observable events during the period. Such events may include a successful (or failed) reception of a bit (or packet), the result of a channel contention, and the end-to-end delay of a message. We denote an event as $\mathbf{O}_k$.

These events reveal different states such as SINR of a link, channel access probability, and quality of a route, but in a probabilistic way. That is, an event $\mathbf{O}_k = \mathbf{o}_k$ happens with a probability for a given network state $\mathbf{V} = \mathbf{v}$, i.e.,

$$\mathbb{P}\{\mathbf{O}_k = \mathbf{o}_k | \mathbf{V} = \mathbf{v}\} = f(\mathbf{o}_k, \mathbf{v}). \quad (3)$$

The function $f(\mathbf{o}_k, \mathbf{v})$ is often available, either mathematically (e.g., the bit error probability for a given SINR), or through training.

Then, if a sequence of independent events $\{\mathbf{o}_k\}_{k=1,\dots,K}$ have happened, we have

$$\mathbb{P}\{\{\mathbf{O}_k\}_{k=1,\dots,K} = \{\mathbf{o}_k\}_{k=1,\dots,K} | \mathbf{V} = \mathbf{v}\} = \prod_{k=1}^{K} f(\mathbf{o}_k, \mathbf{v}), \quad (4)$$

Therefore, we have the *a-posteriori* distribution

$$\mathbb{P}\{\mathbf{V} = \mathbf{v} | \{\mathbf{O}_k\} = \{\mathbf{o}_k\}\} = = \frac{\mathbb{P}\{\{\mathbf{O}_k\} = \{\mathbf{o}_k\} | \mathbf{V} = \mathbf{v}\}}{\int_{\mathbf{v}} \mathbb{P}\{\{\mathbf{O}_k\} = \{\mathbf{o}_k\} | \mathbf{V} = \mathbf{v}\} \cdot \mathbb{P}\{\mathbf{V} = \mathbf{v}\}\, \mathrm{d}\mathbf{v}} \cdot \mathbb{P}\{\mathbf{V} = \mathbf{v}\}, \quad (5)$$

where $\mathbb{P}\{\mathbf{V} = \mathbf{v}\}$ is some a-priori distribution of $\mathbf{V}$, which represents the ideal network state (i.e., without attacks). Note that this quantity is usually available. For example, the distribution of SINR on a link can be derived from the fading model, the channel access probability for any node in a network running CSMA/CA is approximately the same, and so forth. If the accurate knowledge is not available, usually it is still possible to know some information on it, such as the functional form, and the range for the values, as explained in [17].

With the a-posteriori distribution in (5) available, the defender is now aware of the attack activities of the attacker. However, we still need to estimate the attack objective, since it is possibly on another layer than that with the attack activities. For this we employ a classifier taking both the a-posteriori distribution of the network state, and features of attacks that the network is prone to, as input. We use $B$ to denote the features of an alleged attack, which represents a-priori knowledge on typical network state under this attack. Then we have the classifier

$$C(\mathbb{P}\{\mathbf{V} = \mathbf{v} | \{\mathbf{O}_k\} = \{\mathbf{o}_k\}\}, B) \underset{\text{Not attacked}}{\overset{\text{Attacked}}{\lessgtr}} C_{\text{th}}, \quad (6)$$

where $C_{\text{th}}$ is a threshold that should be set according to the typical impact of the alleged attack on network state.

### C. Attack Mitigation

The attack mitigation engine works when an attack is detected. Since the strategy of the attacker may result in a network state in which the optimization in (2) increasing both performance and risk, the mitigation engine must decide a strategy $\mathbf{S}_k$ to optimize the security-performance trade-off. To achieve this, we define a security-performance function as

$$\begin{aligned} h(\mathbf{S}_k, \mathbf{A}_k) &= \alpha \cdot E[U(\mathbf{V}_k)] - \beta \cdot E[G(\mathbf{V}_k)] \\ &= \alpha \cdot E[U(g(\mathbf{S}_k, \mathbf{A}_k))] - \beta \cdot E[G(g(\mathbf{S}_k, \mathbf{A}_k))], \end{aligned} \tag{7}$$

where $E[\cdot]$ is the expectation of a random variable. The expected utility $E[U(\mathbf{V}_k)]$ represents performance and negative attack gain $-E[G(\mathbf{V}_k)]$ represents security. The mitigation engine decides the optimal strategy by solving

$$\underset{\mathbf{S}_k}{\text{maximize}} \quad h(\mathbf{S}_k, \mathbf{A}_k) \tag{8}$$

$$\text{subject to} \quad E[U(\mathbf{V}_k)] \leq U_{\text{th}} \tag{9}$$

$$E[G(\mathbf{V}_k)] \leq G_{\text{th}}, \tag{10}$$

where $\alpha$ and $\beta$ are control variables. By adjusting their values, the controller can regulate the desired trade-off point between performance and security. Constraints (9) and (10) represent the minimum desired performance and security, respectively.

The values of $\alpha$ and $\beta$ will mainly depend on the considered application. For example, video streaming applications usually require high performance on data rate but low security level, whereas a wireless sensor network collecting scientific data usually requires high security level but may tolerate low data rate.

## III. Use-case Example: Hammer-and-anvil Attack

To illustrate how the proposed framework works in practical scenarios, we apply it to a specific cross-layer attack, namely the hammer-and-anvil attack introduced in [5]. Readers can refer to [5] for details.

### A. Brief Introduction

As shown in Figure 1, the considered wireless multi-hop network consists of a set $\mathcal{N}$ of nodes. Data is generated at source nodes and forwarded to the sink $z \in \mathcal{N}$ in a hop-by-hop manner. A node $n \in \mathcal{N}$ must decide its power allocation $\mathbf{P}_n = \{P_n^1, \ldots, P_n^{|\mathcal{F}|}\}$ on the set $\mathcal{F}$ of channels, as well as the next hop $\hat{m}_n$. Therefore, its strategy can be written as $\mathbf{S}_n = \{\mathbf{P}_n, \hat{m}_n\}$.

The objective of node $n \in \mathcal{N}$ is to minimize the expected end-to-end delay from itself to the sink, represented as $T_n(\mathbf{P}_n, \hat{m}_n)$. To this end, a joint optimization of $\mathbf{P}_n$ on physical layer, and $\hat{m}_n$ on network layer is performed, with the optimal strategy

$$\{\mathbf{P}_n^*, \hat{m}_n^*\} = \arg \max_{\mathbf{P}_n \in \mathcal{P}_n, \hat{m} \in \mathcal{V}_n} -T_n(\mathbf{P}_n, \hat{m}_n), \tag{11}$$

where $\mathcal{P}_n$ and $\mathcal{V}_n$ are the set of possible power allocation, and the neighbor set of $n$, respectively.

As a remark, the queuing model in [5] states that the average delay of link $n \to \hat{m}_n$ is an decreasing function of the achievable throughput of the link, which is expressed as

$$\mu_n(\hat{m}_n) = \sum_{f \in \mathcal{F}} \log_2 \left( 1 + \frac{P_n^f H_n^f(\hat{m}_n)}{I_n^f(\hat{m}_n) + \eta_n^f(\hat{m}_n)} \right), \tag{12}$$

with $H_n^f(\hat{m}_n)$, $I_n^f(\hat{m}_n)$, and $\eta_n^f(\hat{m}_n)$ denoting the channel coefficient of link $n \to \hat{m}_n$, the interference and noise at $\hat{m}_n$, respectively.

The expression in (12) establishes the ground for the hammer-and-anvil attack to work. The attacker is made up by a jammer $j$ and a compromised node $c \in \mathcal{N}$. The compromised node can undermine communication in various ways. For the sake of generality, we assume the activities are not observable by an outsider, meaning there is no way to distinguish it from legitimate nodes.

The jammer aims to redirect traffic to the compromised node. To this end, it selectively jams links that do not lead to the compromised node. For example, if $j$ jams the link $n \to m$, the achievable throughput $\mu_n(m)$ will be degraded, resulting in increased $T_n(m)$. With appropriate jamming, the utility is degraded so that, for $\forall \mathbf{P}_n \in \mathcal{P}_n$, $-T_n(\mathbf{P}_n, m) < -T_n(\mathbf{P}'_n, l)$, for $l$ and some $\mathbf{P}'_n$. Then $n$ will choose $l$ instead of $m$ as the next hop.

The strategy of the attacker is simply the power allocation of the jammer, $\mathbf{P}_j$, and its attack gain is the input data rate of the compromised node.

### B. Attack Detection

It is shown in [5] that the attack objective can be achieved by small-scale jamming. Therefore, it is difficult to accurately detect the attack using traditional jamming detection methods based on packet delivery ratio (PDR). We will adopt the detection method in the framework to address this challenge.

In the following, we will focus on link $n \to m$, and thus eliminate the subscripts $n$ and $m$. Moreover, since all the channels are i.i.d., we will also eliminate $f$. We still retain the subscript $j$ to denote the power and channel coefficients involving the jammer $j$. Therefore, we use $P, P_j$ to denote the power allocation of $n$ and $j$ on $f$. Furthermore, $H, H_j$ represent the channel coefficients from $n$ and $j$ to $m$, on channel $f$, respectively.

To detect the attack, we focus on the network state of the interference $I = P_j H_j$, as it is the direct variable affected by jamming. For the observable events, we choose the reception of bits at the receiver, since it reveals the interference. We denote an event as $e_k$ and formally defined it as

$$e_k = \begin{cases} 1, & \text{bit } k \text{ is received correctly,} \\ 0, & \text{bit } k \text{ is not received correctly.} \end{cases} \tag{13}$$

The conditional probability for an event $e_k$ given an interference value, shown as

$$\mathbb{P}\{e_k | I = i\} = \mathbb{P}\{e_k | \gamma = \frac{PH}{i + \eta}\}, \tag{14}$$

is essentially the bit error probability given the SINR $\gamma$, with Gaussian noise $\eta$. (14) is readily available for additive white Gaussian noise (AWGN) channels, and can be obtained through training in other environments.

Let us assume a sequence of $K$ bits is transmitted during a strategy updating period. Therefore we have a sequence of events $\{e_k\}_{k=1,\ldots,K}$ and

$$\mathbb{P}\{\{e_k\}_{k=1,\ldots,K}|I=i\} = \prod_{k=1}^{K} \mathbb{P}\{e_k|I=i\}. \tag{15}$$

Thus, the a-posteriori distribution of $I$ can be estimated as

$$\mathbb{P}\{I=i|\{e_k\}_{k=1,\ldots,K}\} = \frac{\prod_{k=1}^{K} \mathbb{P}\{e_k|I=i\}}{\int_j \prod_{k=1}^{K} \mathbb{P}\{e_k|I=j\}\mathbb{P}\{I=j\}\mathbf{d}j} \cdot \mathbb{P}\{I=i\}. \tag{16}$$

The expression in (16) provides a procedure to estimate the interference of link $n \to m$ on channel $f$. Starting with some a-priori distribution $\mathbb{P}\{I=i\}$, once a sequence of bits has been received (correctly or incorrectly) on channel $f$, the receiver computes the a-posteriori distribution according to the corresponding likelihood for events $\{e_k\}_{k=1,\ldots,K}$. This is done recursively for every transmission. For practical reasons, the integration in (16) is approximated by a summation over a finite set $\mathcal{I}$. Then, it can be rewritten as

$$\mathbb{P}\{I=i|\{e_k\}_{k=1,\ldots,K}\} = \frac{\prod_{k=1}^{K} \mathbb{P}\{e_k|I=i\}}{\sum_{j\in\mathcal{I}} \prod_{k=1}^{K} \mathbb{P}\{e_k|I=j\}\mathbb{P}\{I=j\}} \cdot \mathbb{P}\{I=i\}. \tag{17}$$

When the number of channels is large, it may not be practical to compute (17) for each channel in every transmission. In this case, the receiver can select some channel(s) from $\mathcal{F}$ to update.

For attack mapping, without observable behavior of the compromised node, the most obvious feature is the small jamming scale. We design the classifier as

$$\mathbb{P}\{I_{\text{lwr}} \leq I \leq I_{\text{upp}}\} \quad \underset{\text{attacked}}{\overset{\text{unattacked}}{\lesseqgtr}} \quad \mathbb{P}_{\text{th}}. \tag{18}$$

If the a-posteriori probability that the interference is between a range is higher than a probability threshold, the link is considered to be jammed. The thresholds $I_{\text{lwr}}$, $I_{\text{upp}}$, and $\mathbb{P}_{\text{th}}$ correspond to the feature $B$ in (6). The network management entity can adjust these thresholds according to its information on whether the attack exists, how far the jammer is likely to be, and so forth. The resulting jamming detection scheme is described in Algorithm 1.

**Algorithm 1** Bayesian learning-based jamming detection

```
Given node m ∈ N and a-priori distribution P{I_m^f = i};
while true do
    if A node n ∈ N selects m as the receiver and wins the channel competition then
        n and m conduct channel estimation and get H_nm^f, ∀f ∈ F;
        Based on H_nm^f, n decides the strategy and transmits;
        m selects a F' = {f ⊂ F : P_n^f ≠ 0};
        for f ∈ F' do
            Compute the likelihood in (15) based on the bit reception events;
            Update the a-posteriori probability in (17);
        end for
    end if
    if A-posteriori distribution satisfies convergence condition; then
        Break;
    end if
end while
if (18) holds for a certain number of f ∈ F then
    Node m is considered jammed;
else
    Node m is considered not jammed;
end if
```

### C. Attack Mitigation

After the attack is detected, we further apply the proposed framework to generate a mitigation scheme. Therefore, we define a performance-security function as in (7) and optimize it in the fashion of (8) - (10).

The performance component is simply the negative of delay $-T_n$. The security component is the negative of the attack gain, i.e., the amount of data that is routed to the compromised node $c$. However, since the compromised node is indistinguishable from legitimate nodes, we cannot detect its accurate location or even assert that such a node exists. Therefore, we cannot exactly quantify the attack gain. Instead, we define by $R_n$ the approximate risk of leading to a compromised node, and use $-R_n$ to represent the security component. The optimization problem can then be formulated as

$$\text{maximize} \qquad -\alpha T_n - \beta R_n. \tag{19}$$

The values of $T_n$ and $R_n$ will be discussed later, for different strategies. Without loss of generality, we assume that $l$ is the best next hop candidate for $n$ with jamming, while the jammed node $m$ is the second best one, in the following discussion.

Two simple strategies for $n$ are (i) to reroute to $l$ or (ii) to keep routing to $m$. We denote these strategies as $S_l$ and $S_m$, respectively. The delay for the two strategies are then $T_n(S_l) = T_n(l)$ and $T_n(S_m) = T_n(m)$, respectively. It is reasonable to assume that routing through $l$ has a high likelihood of leading to the compromised node, while choosing $m$ is unlikely to lead to the compromised node. Therefore, we can set $R_n(S_l) = 1$ and $R_n(S_m) = \epsilon$, with a small $\epsilon \in (0, 1]$. Since $R_n(S_l) >> R_n(S_m)$, a desirable compromise between performance and security may not be available with these two strategies alone.

A more sophisticated strategy that may enable a better compromise can be designed based on a secure network coding strategy [19], which we denote as $S_{SNC}$. We construct the secure network coding scheme as follows:

1) Choose a suitable integer $r$ and get the message vector

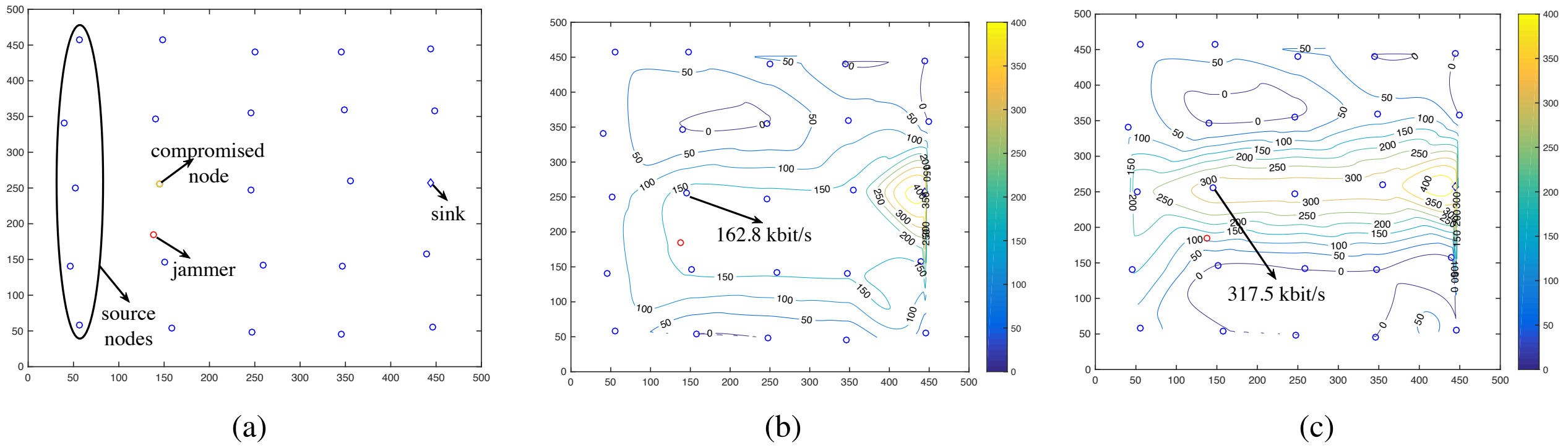

(a) (b) (c)

Fig. 4. Example scenario: (a) topology; (b) traffic map without jamming; (c) traffic map with jamming, $P_j^{\max} = 10\,\mathrm{mW}$.

$X$ from $GF^{(r)}(q)^1$, for some $q$;

2) Choose a suitable $r$-dimensional linear network code $E$ on $GF^{(r)}(q)$;
3) Encode the vector $X$ by multiplying it with the encoding matrix, and get the encoded vector $Y = EX$;
4) Among the $r$ elements of the encoded vector $Y$, the first $N_l$ are transmitted through $l$, and the remaining $N_m$ are transmitted through $m$, $(N_l + N_m = r)$.

According to Theorem 2 in [19], there exists a code $E$ guaranteeing that neither $l$ nor $m$ can decode the messages if $\dim(E) = r$ and $\max(\dim(V_l), \dim(V_m)) < r$,[2] with $V_l$ and $V_m$ being the linear spans of the encoding vectors corresponding to $l$ and $m$, respectively.

Since $N_l$ and $N_m$ out of $N_l + N_m$ encoded messages are transmitted through $l$ and $m$ respectively, the achievable performance for this strategy is

$$T_n(S_{SNC}) = \frac{N_l T_n(l) + N_m T_n(m)}{N_l + N_m} + \frac{1}{\lambda_n}(N_l + N_m - 1). \tag{20}$$

The second term on the RHS represents the additional delay introduced by waiting for all the $N_l + N_m$ messages to arrive to the destination before encoding. Since it is guaranteed that neither $m$ nor $l$ can decode the message, the risk is $R_n(S_{SNC}) = 0$.

To sum up, node $n$ chooses a strategy among (i) reroute to $l$, (ii) keep sending through $m$, and (iii) use secure network coding, according to the performance risk values for each of them. The corresponding performance-security functions are

$$h_n(S_l) = -\alpha \hat{T}_n(S_l) - \beta, \tag{21}$$

$$h_n(S_m) = -\alpha \hat{T}_n(S_m) - \beta\epsilon, \tag{22}$$

$$h_n(S_{SNC}) = -\alpha \hat{T}_n(S_{SNC}). \tag{23}$$

Note that the performance is normalized to be comparable to the risk, which takes values in $[0, 1]$. Then, the problem is

$$\underset{S \in \{S_l, S_m, S_{SNC}\}}{\text{maximize}} \quad h_n(S). \tag{24}$$

This problem can be solved using Algorithm 2.

[1] $GF$ stands for Galois Field.
[2] dim stands for dimension.

**Algorithm 2** Secure network coding based attack mitigation

Given a jammed link $n \to m$:
$n$ finds the best next hop candidate $l \in \mathcal{V}_n$, and the optimal secure network code $(N_l, N_m)$;
**if** $(23) \geq \max[(21), (22)]$ **then**
  proceed with the secure network coding scheme;
**else if** $(22) \geq \max[(21), (23)]$ **then**
  Transmit through $m$;
**else**
  Transmit through $l$;
**end if**

## IV. Performance Evaluation

We now evaluate the proposed schemes through both simulations and testbed experiments.

### A. Simulation Settings

A network of 25 nodes is randomly generated in a $500\,\mathrm{m} \times 500\mathrm{m}$ area, with one sink and one compromised node. There is a jammer whose location is set in accordance to the location of the compromised node. A typical example is shown in Figure 4(a). For convenience, we label the nodes with numbers. Data sessions are generated at the leftmost 5 nodes. The generation rates are randomly set, with a mean value of $80\,\mathrm{kbits/s}$. Each legitimate node has power budget $P^{\max} = 1$ W. There are 10 mutually orthogonal channels, with bandwidth of $10\,\mathrm{kHz}$ each. All channels have a path loss with exponent 3 and i.i.d. Rayleigh fading with parameter 0.5. The spectral density of noise is $1 \times 10^{-8}$ W/Hz. The power budget of the jammer is set to $10 - 100\,\mathrm{mW}$, i.e., only $1/100 - 1/10$ as high as that of the legitimate nodes. Even with this level of jamming power, the attack is effective. This can be verified in Fig. 4 (b) and (c), which show traffic maps corresponding to the example topology. Considering the simulation settings, even with a jammer that only causes interference around 10 times the noise level, a significant amount of data can be "driven" to the compromised node. Therefore, the efficiency and stealthiness is verified.

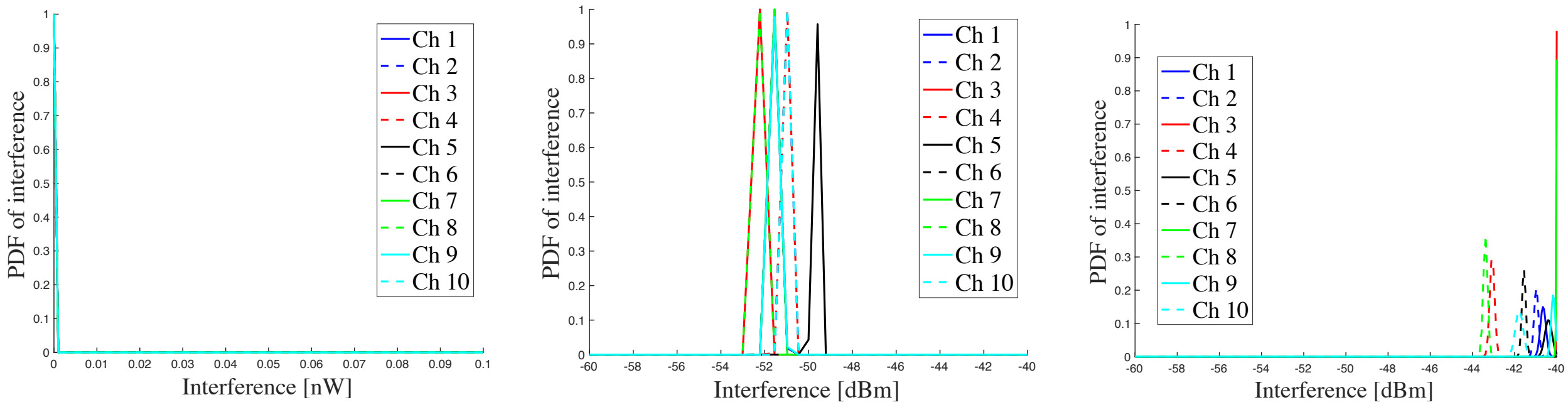

(a) (b) (c)

Fig. 5. Detection results for small-scale jamming: (a) $P_j^{\max} = 0$; (b) $P_j^{\max} = 10$ mW; (c) $P_j^{\max} = 100$ mW.

| Jamming power budget $P_j^{\max}$ (mW) | 0 | 10 | 100 |
|---|---|---|---|
| Bit error rate (%) | 3.84 | 9.84 | 9.78 |

TABLE I

BER OF NODE 7 FOR DIFFERENT JAMMING POWER

### *B. Attack Detection*

Since we focus on small-scale jamming, the considered interference range for the distribution $\mathcal{I}$ is set to $[0, 1 \times 10^{-7}]$W, with a step size of $1 \times 10^{-9}$ W. The proposed jamming detection scheme can effectively detect the interference caused by small-scale jamming attack. For node 7, which is the target node of the jammer, the a-posteriori distribution of $I$ is shown in Figure 5, for $P_j^{\max} = 0$, $P_j^{\max} = 10$ mW, and $P_j^{\max} = 100$ mW. We assume uniform a-priori distribution $\mathbb{P}\{I = i\}$. We observe that the interference can be effectively and accurately estimated in all cases. For $P_j^{\max} = 0$, which means no jamming attack, the distribution converges to a single peak at $I = 0$ for all channels, and provides a perfect estimation. For $P_j^{\max} = 10$ mW, the a-posteriori distribution of $I$ concentrates at around $-51$ dBm. Considering that the distance from the jammer to node 7 is 40.2 m, the average jamming-caused interference is $-50.5$ dBm. Therefore, the result is accurate. For $P_j^{\max} = 100$ mW, the a-posteriori distribution converges to the largest possible value in $\mathcal{I}$, i.e., $I = -40$ dBm, since in this case the jammer creates relatively large interference.

To provide a better understanding, we list the corresponding bit error rates (BER) in Table I. We observe that the difference in BER with or without jamming is small. This implies that a detection scheme based on BER exclusively may not work well. Therefore, the proposed detection scheme can better discriminate between jammed and unjammed scenarios with small-scale jamming. This is confirmed by the receiver operating characteristic (ROC) curves shown in Figure 6.

To obtain the ROC curve, we ran simulations for different topologies and varied the jamming power budget between 10 mW and 100 mW with a step of 10 mW. For each simulation, the learned distribution of interference is fed to the classifier defined in (18). We point out that, there is not yet a detection method for the considered attack to compare with. We try to compare the ability of attack activity detection, i.e., jamming detection, with traditional methods. Therefore, we set the upper bound $I_{\text{upp}}$ to infinity for fairness. True positive rates (TPR) and false positive rates (FPR) are recorded for the threshold $I_{\text{lwr}} \in [0, 1 \times 10^{-4}]$mW, with a step of $1 \times 10^{-6}$mW. For comparison, we also plot the ROC curve for BER-based jamming detection, with the classifier

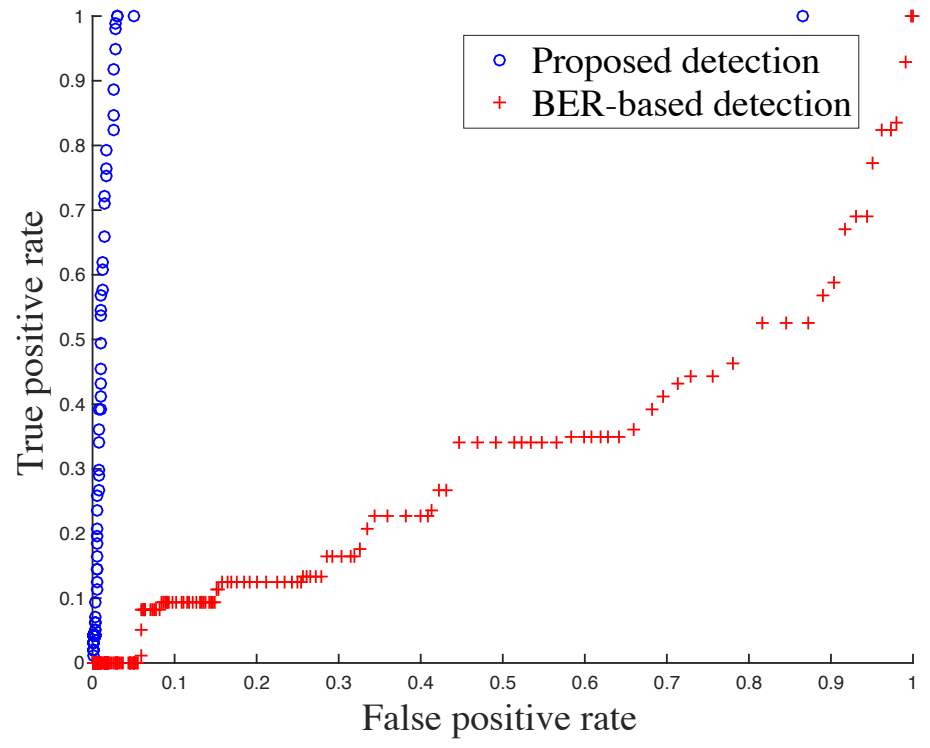

Fig. 6. ROC curves of the proposed detection and BER-based detection methods.

$$BER \underset{\text{jammed}}{\overset{\text{unjammed}}{\lessgtr}} BER_{\text{th}}, \quad (25)$$

and the threshold varying from 0 to 0.5, with a step of 0.001. Comparing the two ROC curves, we find that the proposed detection scheme results in a large area under the curve (AUC), and can thus achieve very reliable detection results in most cases. Conversely, the BER-based detection method has much smaller AUC, and thus is inferior to the proposed method.

### *C. Attack Mitigation*

To demonstrate the performance of the mitigation scheme, we used traces from the previous simulations. We identified the jammed link in each topology, i.e., $n \to m$, and found the best next hop candidate $l$ for $n$. For the strategies of rerouting to $l$, staying on $m$, and using secure network coding, we computed the performance-risk functions defined in (21), (22) and (23), for different values of $(\alpha, \beta)$. The results are shown in Figure 7.

For $\alpha = 1, \beta = 0$, i.e., when performance is the only concern, secure network coding is not as good as using the best unjammed link. However, it still achieves considerable gain compared to the jammed link. When risk is taken into account, the benefits of secure network coding become obvious. For $\alpha = \beta = 0.5$, i.e., when performance and risk are equally important, secure network coding outperforms the other two since it provides a better performance-risk balance. For $\alpha = 0, \beta = 1$, i.e., when the only concern is risk, the relationship between the three strategies is the same as in the previous case, but the benefits of secure network coding compared to using the jammed link become marginal, since it is unlikely that a link leading to a compromised node is jammed.

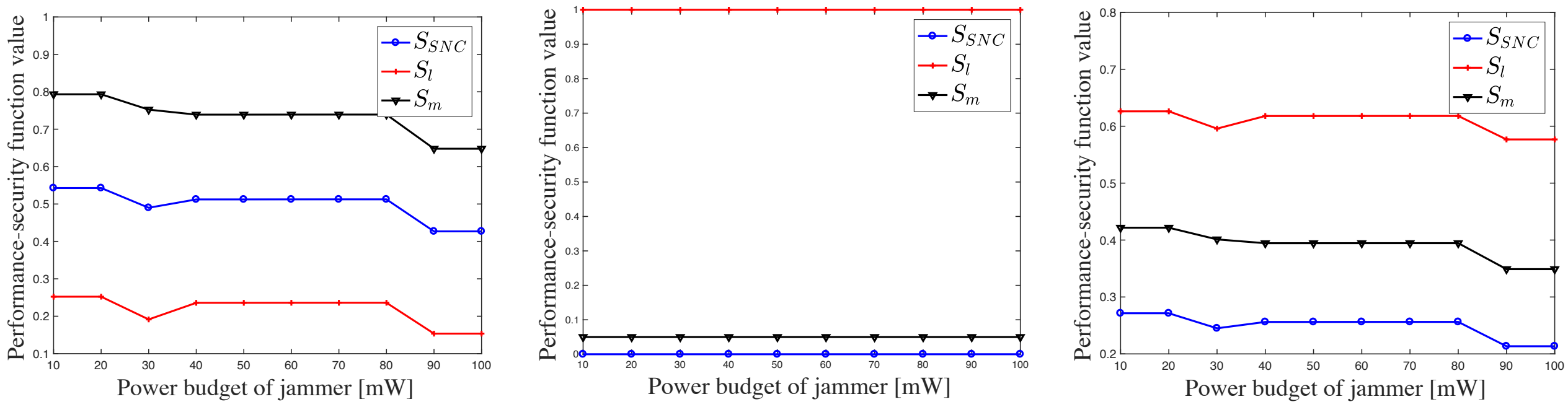

(a) (b) (c)

Fig. 7. Optimal performance-security trade-off with different settings: (a) $\alpha = 1, \beta = 0$; (b) $\alpha = 0, \beta = 1$; (c) $\alpha = 0.5, \beta = 0.5$.

### D. Testbed Evaluation

In real scenarios, there may not be a closed-form relationship between BER and SINR. In this case, a training process is required. We conducted test-bed experiments where we first trained a BER curve from the transmission results without jamming; then, based on the curve, we ran the jamming detection scheme with reactive jamming to show its accuracy. We used two USRP N210 [18] to form a transmitter-receiver pair, and another USRP X310 [18] is used as the jammer, emitting interference. We used a channel with bandwidth of 10 kHz.

In the training period, the jammer constantly emits interference with varying power. Meanwhile, the transmitter-receiver pair conducts normal transmissions. The receiver estimates the SNR and calculates the corresponding BER by comparing the received file with the original one. Multiple points are obtained, based on which a BER-SINR curve is generated by using exponential fitting. The curve is shown in Figure 8.

To test the accuracy of the proposed detection, we run more experiments with the reactive jamming model. The receiver is able to estimate the SINR without jamming. Along with the real bit reception events, the distribution of interference can be learned. We use the interference-to-noise ratio (INR) instead of absolute interference level to reduce the error. In Figure 9, we show the resulting PDF for different jamming scenarios. Since the real interference level at the receiver is unknown, we use the BER to distinguish the scenarios. Note that there is an irreducible BER caused by noise and the multi-path effect, representing the lowest BER the transmitter-receiver pair is able to achieve.

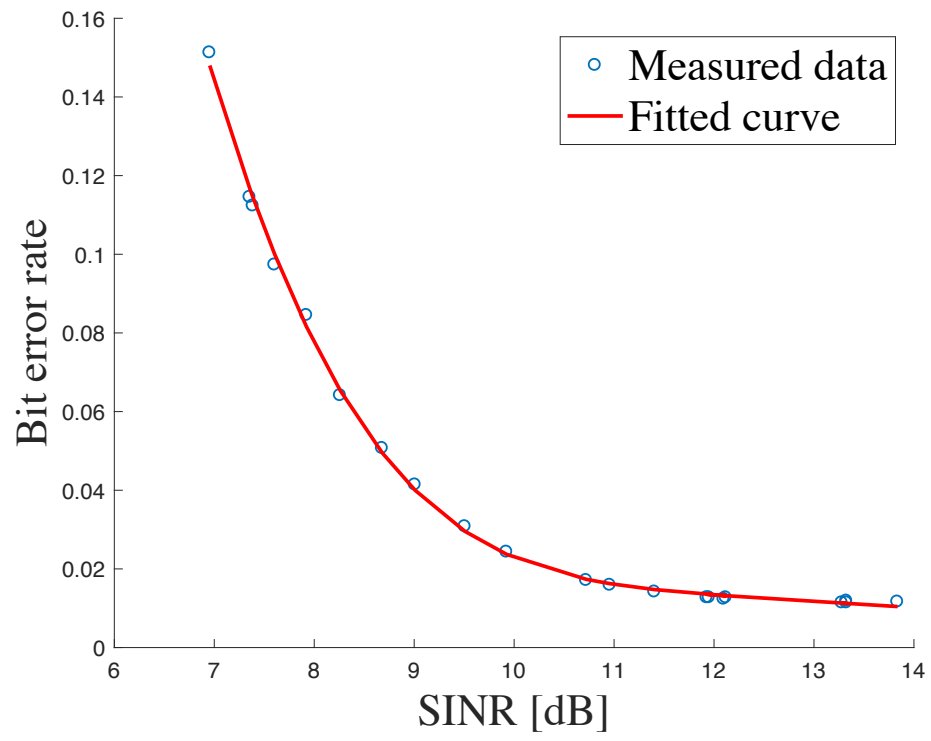

Fig. 8. Trained BER curve.

| Detection Result | No | Yes | Yes |
|---|---|---|---|
| Relative BER | 0.27% | 1.28% | 3.92% |

TABLE II
DETECTION RESULT AND CORRESPONDING RELATIVE BER

Since our objective is to detect small-scale jamming, we set $INR_{\text{th}} = 0$ dB, and used $Pb(INR > INR_{\text{th}}) > 0.9$ as the classifier. If the interference is at least at the same level of noise with probability 0.9, then it is considered to be jammed. With this classifier, the detection results and the relative BER (real BER - irreducible BER) are shown in Table II. The proposed scheme is able to detect small scale jamming resulting in very low relative BER, as long as it is greater than 1%, which verifies the effectiveness of the detection scheme.

## V. Related Work

Recently, several cross-layer attacks have been unveiled, which cover all layers of the network protocol stack. In [7], two cross-layer attacks jointly utilizing tools at the physical, MAC, and network layers are studied. Vulnerabilities of TCP have been exploited by researchers in multiple ways. In [6], a primary emulation attack is used with the same objective. The hammer-and-anvil attack [5] uses jamming to assist the objective of routing manipulation. Although many cross-layer attacks have been proposed, little work has been devoted to devise general countermeasures to address them, as opposed to this paper. There are some efforts on cross-layer intrusion detection [20], [21]. While these works support using features

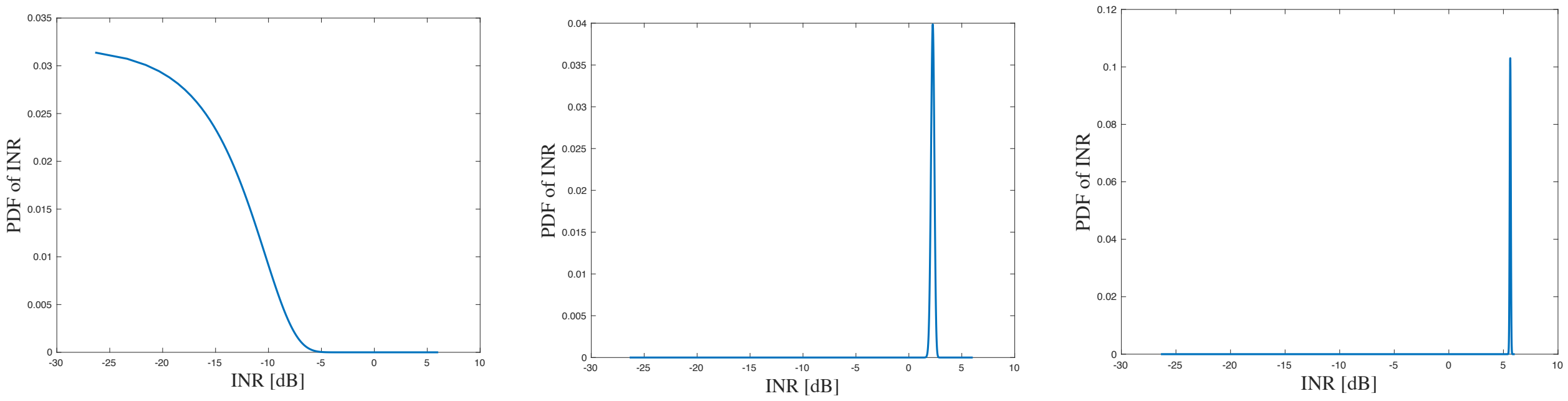

(a) (b) (c)

Fig. 9. Detection results for small-scale jamming, with real BER: (a) 1.42%; (b) 2.43%; (c) 5.07%.

on multiple layers to detect an attack, these features are not considered in a cross-layer manner.

There is a rich body of literature on jamming detection in wireless networks [12], [13], [22], [23]. Most of them use hard thresholds on metrics such as packet delivery ratio (PDR) for hypothesis making. Some works proposed to use optimal likelihood ratio (LLR) test and sequential probability ratio test (SPRT), but perfect knowledge on jamming effect is still needed, which is generally not available for small-scale jamming.

## VI. Conclusions

We proposed a general countermeasure framework against cross-layer attacks in wireless networks, with a learning-based detection and optimal mitigation scheme. The learning-based detection is able to utilize observations of multiple layers to generate a hypothesis on the risk of a certain attack; the mitigation scheme optimally uses multiple mitigation tools and achieves the desired security-performance trade-off under different requirements. We tested the framework on a state-of-the-art cross-layer attack with implementation details. Both simulation and test-bed results were provided, showing that the countermeasure is superior to traditional single layer approaches.